\documentclass[aps,preprint]{revtex4}
\usepackage{amsfonts}
\usepackage{amsmath}
\usepackage{amssymb}
\usepackage{graphicx}

\input{tcilatex}

\begin{document}

\title{Multiple quantum dynamics in dipole-coupling spins in solid.}
\author{G.B. Furman, S.D. Goren, V. M. Meerovich, V. L. Sokolovsky, and
S.William }
\affiliation{Physics Department, Ben Gurion University, Beer Sheva, 84105, Israel}

\begin{abstract}
A perturbation method deals with dipolar coupling spins in solids is
presented. As example of application the method, the multile-quantum
coherence dynamics in clusters of a linear chain of four nuclear spins and a
ring of six spins coupled by dipole-dipole interaction are considered. The
calculated 0Q- and 2Q intensities in a linear chain of four nuclear spins
and 6Q intensity in a ring of six spins vs the duration of the preparation
period agree well with the exact solutions (for linear chain of four nuclear
spins) and simulation data (for linear chain of four nuclear spins and a
ring of six spin).
\end{abstract}

\maketitle

\textbf{Introduction}

The dipole-dipole interaction (DDI) is widespread in nature and determines
behavior of many physical systems. In solids state nuclear magnetic
resonance (NMR), it is responsible for many specific phenomena and plays an
important role in nuclear spin dynamics \cite{A.Abragam}. However in solids
the evolution of a spin system under the DDI involves many spins and
behavior of such system cannot be analyzed analytically. Even the numerical
analysis becomes difficult because the number of states $N=2^{n}$ is growing
exponentially with the increasing of number of spins $n$. Therefore existing
theories are based on phenomenological models and only macroscopic
characteristics such as spin-spin relaxation times, the second and the
fourth moments of resonance lines are taken into account \cite{A.Abragam}.
These difficulties are very clearly \ displayed in multiple-quantum (MQ)
spin dynamics. The MQ phenomena involve various multiple-spin transitions
between the Zeeman energy levels and MQ coherence is formed at time $%
t>\omega _{d}^{-1}$ , where $\omega _{d}$ is the characteristic frequency of
DDI \cite{J.B.Murdoch}. Hence, $\omega _{d}t>1$ is not a small parameter,
that did not allow one to use perturbation methods to study MQ dynamics.
Indeed only simple exactly solvable models of a spin system such as two and
three dipolar coupling spins 1/2 \cite%
{A.K.Roy&K.K.Gleason1996,S.I.Doronin2003} or one-dimensional linear chains
of spins \cite{S.I.Doronin2000} were analyzed theoretically. The last
achievement in this direction is the model with identical DDI coupling
constants for all spin pairs \cite{J.Baugh2001,M.G.Rudavets&E.B.Fel'dman2002}%
. Note that the simplified calculations essential for the case of identical
DDI coupling constant have been already mentioned \cite{Lowe}. This model
describes only zero- and second-order coherences. But in real solids, DDI is
characterized by the different coupling constants. In order to obtain
important information on molecular structure and spin dynamics of these
systems, the evolution of large number modes of MQ coherence has to be
analyzed.

Another important purpose of the study of MQ coherence is the field of
quantum information processing \cite{D.G.Cory1997,N.Gershenfeld,B.Kane1998}.
It was experimentally demonstrated the possibility of creating pseudopure
spin states in clusters of coupled spins by MQ method \cite{Khitrin} and
that dynamics of the quantum entanglement is uniquely determined by the time
evolution of MQ coherences \cite{S.I.Doronin2003}.

Our objective is to adapt a perturbation approach to the problem of dipolar
coupling spin dynamics in solids and to obtain the description of large
number modes of MQ coherence. The Hamiltonian of the spin system with
different DDI constants is divided into several parts, each of them is
characterized by its constant. Our main idea is to take into account in MQ
dynamics the influence of the different items of the Hamiltonian with
different degree of accuracy. The items corresponding to smaller DDI
coupling constants are considered as a perturbation while the item with the
largest constant (describing the interaction between nearest neighbours) is
taken into account exactly. The proposed approach will be a power method to
describe wide range of physical problems dealing with dynamics of dipolar
coupling spins in solids. On the one hand, this approach uses the advantages
of exactly solvable models \cite%
{M.G.Rudavets&E.B.Fel'dman2002,A.R.Kessel2002}. On the other hand, it
simplifies calculations by using a perturbation technique.

\textbf{Theory}

Let us consider a system of nuclear spins $1/2$ placed in an external
magnetic field $H_{0}\Vert $ $z$ axis and subjected to a suitable designed
pulse sequence, thus permitting the observation of MQ coherences \cite%
{J.Baum}. The effect of the sequence of irradiating pulses on the spin
system can be represented by a unitary transformation propagator \cite%
{J.Baum} 
\begin{equation}
U\left( t\right) =e^{-it\mathcal{H}},  \tag{1}
\end{equation}%
where $\mathcal{H}$ is the effective time-independent Hamiltonian. First
assume that the effective Hamiltonian can be divided into two parts with
only two different DDI constants $\alpha $ and $\beta $: $\mathcal{H}=%
\mathcal{A}+\mathcal{B}$ , where the operators $\mathcal{A}$\ and $\mathcal{B%
}$ can be presented in the following form: $\mathcal{A}=\alpha A$ and $%
\mathcal{B}=\beta B$, $\alpha =\left\| \mathcal{A}\right\| $ and $\beta
=\left\| \mathcal{B}\right\| $ are norms of operators $\mathcal{A}$ and $%
\mathcal{B}$, respectively ($\alpha >$ $\beta $ and $\left[ A,B\right] \neq
0 $). The operators $A\ $\ and $B$ do not include any DDI coupling
constants. We present expression (1) in \ the form

\begin{equation}
e^{-it\left( \alpha A+\beta B\right) }=e^{-it\beta B}\sigma _{A}\left(
t\right) ,  \tag{2}
\end{equation}%
where operator $\sigma \left( t\right) $ obeys the differential equation

\begin{equation}
i\frac{d\sigma \left( t\right) }{dt}=\alpha A\left( t\right) \sigma
_{A}\left( t\right) ,  \tag{3}
\end{equation}%
with the initial condition 
\begin{equation}
\sigma _{A}\left( 0\right) =1  \tag{4}
\end{equation}%
and $A\left( t\right) =e^{-it\beta B}Ae^{it\beta B}$. Assume that $\alpha >$ 
$\beta $. Solving (3), we limit ourselves by the first power of $\frac{\beta 
}{\alpha }$. Expanding the exponential term in the right hand of Eq. (2) in
a series and substituting in Eq. (3), we obtain

\begin{equation}
i\frac{d\sigma _{A}\left( t\right) }{dt}=\alpha \left( A-it\beta \left[ B,A%
\right] \right) \sigma _{A}\left( t\right) .  \tag{5}
\end{equation}%
To solve Eq.(5) we will use the iterative method. We will search the
solution of Eq.(5) as a series on parameter $\left( \alpha t\right) ^{n}$ $:$

\begin{equation}
\sigma _{A}\left( t\right) =\sum_{n=0}^{\infty }\sigma _{A}^{\left( n\right)
}\left( t\right)  \tag{6}
\end{equation}

Taking into account that $\sigma _{A}^{\left( 0\right) }\left( t\right) =1$
the solution of Eq.(5) for $n=1$ can be obtained:$\sigma _{A}^{\left(
1\right) }\left( t\right) =\left( yA-\frac{1}{2}y^{2}\varepsilon \left[ B,A%
\right] \right) $

\begin{equation}
\sigma _{A}^{\left( 1\right) }\left( t\right) =\left( -i\alpha t\right) A. 
\tag{7}
\end{equation}%
For $n=2$, ones obtains 
\begin{equation}
\sigma _{A}^{\left( 2\right) }\left( t\right) =\frac{\left( -i\alpha
t\right) ^{2}}{2}\left( A^{2}+\frac{\beta }{\alpha }\left[ B,A\right] \right)
\tag{8}
\end{equation}%
Keeping only linear in $\frac{\beta }{\alpha }$ terms the following
expression for operator $\sigma _{A}^{\left( m\right) }\left( t\right) $ can
be obtained

\begin{equation}
\sigma _{A}^{\left( m\right) }\left( y\right) =\frac{\left( -i\alpha
t\right) ^{m}}{m!}\left( A^{m}-\frac{\beta }{\alpha }\left( \left(
m-1\right) BA^{m-1}-\sum_{j=0}^{m-2}A^{m-1-j}BA^{j}\right) \right)  \tag{9}
\end{equation}

Utilizing (9) to expand the expression of the unitary transformation
propagator (1) we find 
\begin{equation}
U\left( t\right) =e^{-it\left( \alpha A+\beta B\right) }=\left( 1-\frac{%
\beta }{\alpha }\sum_{n=0}^{\infty }\frac{\left( i\alpha t\right) ^{n+1}}{%
\left( n+1\right) }\sum_{j=0}^{n}\frac{\left( -1\right) ^{j}}{j!\left(
n-j\right) !}A^{j}BA^{n-j}\right) e^{-i\alpha tA}\text{.}  \tag{10}
\end{equation}%
Taking into account that 
\begin{equation}
\sum_{j=0}^{n}\frac{\left( -1\right) ^{j}}{j!\left( n-j\right) !}%
A^{j}BA^{n-j}=\frac{1}{n!}\left\{ B,A^{n}\right\}   \tag{11}
\end{equation}%
where the repeated commutators bracket $\left\{ ...\right\} $ is defined by 
\begin{equation}
\left\{ B,A^{0}\right\} =B,\text{ \ \ \ \ \ \ \ \ \ }\left\{
B,A^{n+1}\right\} =\left[ \left\{ B,A^{n}\right\} ,A\right] ,  \tag{12}
\end{equation}%
we obtained that expression of the unitary transformation propagator takes a
form 
\begin{equation}
U\left( t\right) =e^{-it\left( \alpha A+\beta B\right) }=\left( 1-\frac{%
\beta }{\alpha }\sum_{n=0}^{\infty }\frac{\left( i\alpha t\right) ^{n+1}}{%
\left( n+1\right) !}\left\{ B,A^{n}\right\} \right) e^{-i\alpha tA}. 
\tag{13}
\end{equation}%
The summing over $n$ in (13) results in 
\begin{equation}
\sum_{n=0}^{\infty }\frac{\left( i\alpha t\right) ^{n+1}}{\left( n+1\right) !%
}\left\{ B,A^{n}\right\} =i\alpha \int dte^{i\alpha tA}Be^{-i\alpha tA}. 
\tag{14}
\end{equation}%
Substituting Eq. (14) into (13) we obtain

\begin{equation}
e^{-it(\alpha A+\beta B)}=\left( 1-i\beta
\int_{0}^{t}dxe^{-ixA}Be^{ixA}\right) e^{-i\alpha tA},  \tag{15}
\end{equation}%
which is a well-known formula for expansion of an exponential operator in a
perturbation series \cite{Bellman,R.M.Wilcox1967}. Expanding the operator
exponent $e^{-i\alpha tA}$ of the left-hand of Eq. (13) gives

\begin{equation}
e^{-it\left( \alpha A+\beta B\right) }=\sum_{k=0}^{\infty }\frac{\left(
-i\alpha t\right) ^{k}}{k!}\left( A^{k}-\frac{\beta }{\alpha }%
\sum_{n=0}^{\infty }\frac{\left( i\alpha t\right) ^{n+1}}{\left( n+1\right) !%
}\left\{ B,A^{n}\right\} A^{k}\right)   \tag{16}
\end{equation}%
After rearranging of the series (16) we obtain%
\begin{equation}
e^{-it\left( \alpha A+\beta B\right) }=\left[ 1-\frac{\beta }{\alpha }%
\sum_{n=0}^{\infty }\sum_{m=0}^{\infty }\frac{\left( -1\right) ^{m}\left(
i\alpha t\right) ^{n+m+1}}{n!m!\left( n+m+1\right) }A^{n}BA^{m}\right]
e^{-i\alpha tA}  \tag{17}
\end{equation}%
In the same way we obtain the expansion up to second order in the ratio $%
\frac{\beta }{\alpha }:$

\begin{align}
e^{-it\left( \alpha A+\beta B\right) }& =\{1-\frac{\beta }{\alpha }%
\sum_{n=0}^{\infty }\sum_{m=0}^{\infty }[\frac{\left( -1\right) ^{m}\left(
i\alpha t\right) ^{n+m+1}}{n!m!\left( n+m+1\right) }A^{n}BA^{m}  \notag \\
& -\left( \frac{\beta }{\alpha }\right) \sum_{l=0}^{\infty
}\sum_{k=0}^{\infty }\frac{\left( -1\right) ^{m+k}\left( i\alpha t\right)
^{k+l+m+n+2}A^{l+m}BA^{k}BA^{l}}{n!m!k!l!\left( k+l+1\right) \left(
k+l+m+n+2\right) }]\}e^{-i\alpha tA}  \tag{18}
\end{align}

Expression (18) can be generalized to a case when the exponential operator
contains arbitrary number of the non-commutative operators and can be
extended to include various powers of the operators. For example, if the
effective Hamiltonian operator includes three non-commutative operators $%
\mathcal{H}=\mathcal{A}+\mathcal{B+C}=\alpha A+\beta B+\delta C$, where $%
\delta \leq \beta <\alpha $, we can obtain the expansion in series for a
three-spin cluster replacing $B$ by $B+\frac{\delta }{\beta }C$ in Eq. (18) .

Expressions (17) and (18) represent independent expansions in two parameters 
$\frac{\beta }{\alpha }<1$\ and $x=\alpha t$. \ Eqs. (17) and (18) appear to
be complicated but in fact it is quite simple to use, as the following
examples will illustrate.

\textbf{Results and discussion}

Let us consider a cluster which constitutes a linear chain of four nuclear
spins $1/2$ coupled by DDI. The MQ dynamics in the rotating frame is
described by the time-independent average Hamiltonian: 
\begin{equation}
\mathcal{H}=\sum_{j<k}\mathcal{H}_{jk}=-\frac{1}{2}\sum_{j<k}d_{jk}\left(
I_{j}^{+}I_{k}^{+}+I_{j}^{-}I_{k}^{-}\right)  \tag{19}
\end{equation}%
and $I_{j}^{+}$ and $I_{j}^{-}$ are the raising and lowering operators for
spin $j$. The dipolar coupling constant, $d_{jk}$, for any pair of nuclei $j$
and $k$, is given by 
\begin{equation}
d_{jk}=\frac{\gamma ^{2}\hbar }{2r_{jk}^{3}}\left( 1-3\cos ^{2}\theta
_{jk}\right) ,  \tag{20}
\end{equation}%
where $\gamma $ is the gyromagnetic ratio of the nuclei, $r_{jk}$ is the
internuclear spacing, and $\theta _{jk}$ is the angle that the vector \ $%
\vec{r}_{jk}$\ makes\ with the external magnetic field. In the
high-temperature approximation the density matrix at the end of the
preparation period is given by 
\begin{equation}
\rho \left( t\right) =e^{-i\mathcal{H}t}\rho \left( 0\right) e^{i\mathcal{H}%
t}  \tag{21}
\end{equation}%
where $\rho \left( 0\right) $ is the initial density matrix in the
high-temperature approximation 
\begin{equation}
\rho \left( 0\right) =\sum_{j=1}^{4}I_{j}^{z},  \tag{22}
\end{equation}%
$I_{j}^{z}$ is the projection of the angular momentum operator of spin $j$
on the direction of the external field. The average Hamiltonian (19) can be
divided into the three parts according to the number of different coupling
constants $d_{12}>d_{13}>d_{14}$ : 
\begin{equation}
\mathcal{H}=\mathcal{H}_{12}+\mathcal{H}_{13}+\mathcal{H}_{14},  \tag{23}
\end{equation}%
where $d_{12}$ is the coupling constant of the nearest neighbors, $d_{13}$
is the coupling constant of the next nearest neighbors, and $d_{14}$ is the
coupling constant of the next-next-nearest neighbors. The experimentally
observed values are the intensities of multiple-quantum coherences $%
J_{nQ}\left( t\right) $:%
\begin{equation}
J_{nQ}\left( t\right) =\frac{1}{Tr\rho ^{2}\left( 0\right) }\sum_{p,q}\text{ 
}\rho _{pq}^{2}\left( t\right) \text{ for }n=m_{zp}-m_{zq}.  \tag{24}
\end{equation}%
where $m_{zp}$ and $m_{zq}$ are the eigenvalues of the initial density
matrix (22). Using expansion (18) up to second order in $\frac{\beta }{%
\alpha }=\frac{d_{13}}{d_{12}}$ and $\frac{\delta }{\alpha }=\frac{d_{14}}{%
d_{12}}$ and keeping terms up to eighth order in $x=d_{12}t$ , one can
determine the normalized $0$-quantum ($J_{0Q}$) and $2$-quantum ($J_{2Q}$)
intensities:

\begin{align}
J_{0Q}& =1-\frac{3}{2}x^{2}\left[ 1+\frac{1}{3}\left( \frac{\delta }{\alpha }%
\right) ^{2}+\frac{2}{3}\left( \frac{\beta }{\alpha }\right) ^{2}\right]
+x^{4}\left[ \frac{7}{6}+\frac{2}{3}\left( \frac{\delta }{\alpha }\right)
^{2}+2\left( \frac{\beta }{\alpha }\right) ^{2}\right]  \notag \\
& -x^{6}\left[ \frac{2}{5}+\frac{1}{3}\left( \frac{\delta }{\alpha }\right)
^{2}+\frac{4}{3}\left( \frac{\beta }{\alpha }\right) ^{2}\right] +\frac{2}{21%
}x^{8}\left[ \frac{47}{60}+\left( \frac{\delta }{\alpha }\right) ^{2}+\frac{%
22}{5}\left( \frac{\beta }{\alpha }\right) ^{2}\right]  \tag{25}
\end{align}%
and%
\begin{align}
J_{2Q}=& -x^{2}\left[ \frac{3}{4}+\frac{1}{2}\left( \frac{\beta }{\alpha }%
\right) ^{2}\right] +x^{4}\left[ \frac{7}{12}+\frac{1}{3}\left( \frac{\delta 
}{\alpha }\right) ^{2}+\left( \frac{\beta }{\alpha }\right) ^{2}\right] 
\notag \\
& -x^{6}\left[ \frac{1}{5}+\frac{1}{6}\left( \frac{\delta }{\alpha }\right)
^{2}+\frac{2}{3}\left( \frac{\beta }{\alpha }\right) ^{2}\right] +\frac{1}{21%
}x^{8}\left[ \allowbreak \frac{47}{60}+\left( \frac{\delta }{\alpha }\right)
^{2}+\frac{22}{5}\left( \frac{\beta }{\alpha }\right) ^{2}\right] .  \tag{26}
\end{align}%
In the case where the parameter$\ x$\ is not small we have to take exactly
into account of the terms containing $x$\ . By summation of series in (12)
over $n$, $m$, $l$, and $k$\ we obtain exact analytical expressions which
give the time dependence of intensities of $0$\ - quantum coherence

\begin{equation}
J_{0Q}=1-2J_{2Q}  \tag{27}
\end{equation}%
and of 2-quantum coherence 
\begin{eqnarray}
J_{2Q} &=&-\frac{1}{4}+\frac{1}{4}\cos (x\sqrt{5})\cos (x)-\frac{x\sqrt{5}}{%
50}\sin (x\sqrt{5})\cos (x)\left[ \left( \frac{\delta }{a}\right)
^{2}+3\left( \frac{\beta }{a}\right) ^{2}\right]  \notag \\
&&-x^{2}\left\{ \frac{\cos (x\sqrt{5})\cos (x)}{5}\left[ \left( \frac{\beta 
}{a}\right) ^{2}+\frac{3}{4}\left( \frac{\delta }{a}\right) ^{2}\right] +%
\frac{\sqrt{5}}{20}\left( \frac{\delta }{a}\right) ^{2}\sin (x\sqrt{5})\sin
(x)\right\}  \TCItag{28}
\end{eqnarray}%
in which the interaction between the nearest neighbours is taken into
account exactly. Now let us compare the time dependence of \ the MQ
intensities given by (25) - (28) with those obtained using an exact solution
of the evolution equation and using computer simulation. The exact solution
gives the following expressions for $J_{0Q}$

\begin{equation}
J_{0Q}=1-2J_{2Q}  \tag{29}
\end{equation}%
and for $J_{2Q}$

\begin{equation}
J_{2Q}=-\frac{1}{4}+\frac{1}{8}\cos \left( K_{+}x\right) \cos \frac{P_{+}x}{2%
}\cos \frac{P_{-}x}{2}+\frac{1}{8}\cos \left( K_{-}x\right) \cos \left(
Px\right)  \tag{30}
\end{equation}%
where%
\begin{equation}
K_{\pm }=\left( \frac{\delta }{\alpha }\right) \pm 1  \tag{31}
\end{equation}

\begin{equation}
P=\left( \left( \frac{\delta }{\alpha }\right) ^{2}+2\left( \frac{\delta }{%
\alpha }\right) +5+4\left( \frac{\beta }{\alpha }\right) ^{2}\right) ^{\frac{%
1}{2}}  \tag{32}
\end{equation}

\begin{equation}
P_{\pm }=R_{+}\pm R_{-}  \tag{33}
\end{equation}

\begin{equation}
R_{\pm }=\left( \left( \frac{\delta }{\alpha }\right) ^{2}-2\left( \frac{%
\delta }{\alpha }\right) +5\pm 8\frac{\beta }{\alpha }+4\left( \frac{\beta }{%
\alpha }\right) ^{2}\right)  \tag{34}
\end{equation}%
Figs. 1 and 2 show the evolution of the normalized 0Q and 2Q coherences in
the linear chain of four nuclear spins $1/2$ coupled by DDI, where $\frac{%
\beta }{\alpha }=\frac{1}{8}$ , $\frac{\delta }{\alpha }=\frac{1}{27}$ and
at $t=0$ the spin system is in thermal equilibrium determinates by matrix
(16). These figures present also the results of computer simulation
performed using the MATLAB package. One can see that expressions (25) and
(26) gives a good agreement with the exact analytical and numerical results
for $x\leq 1$ , while expressions (27) and (28) agree up to $x=5.$ Note that
our results coincide also with the dependences obtained in paper \cite%
{Cho1996}.

As a second example let us consider the MQ dynamics in a ring of six spins
coupled by dipole-dipole interaction. The ratios of the dipole-dipole
coupling constants are given by $\frac{\beta }{\alpha }=\frac{1}{3\sqrt{3}}$
and $\frac{\delta }{\alpha }=\frac{1}{8},$ where $\alpha $ is the coupling
constant of the nearest neighbors, $\beta $ is the coupling constant of the
next-nearest neighbors, $\delta $ is the coupling constant of the
next-next-nearest neighbors. Calculation using expansion (18) up to second
order in ratios of $\frac{\beta }{\alpha }$ and $\frac{\delta }{\alpha }$
and eighth order in $x=\alpha t$ gives zero intensity for 6Q coherence. The
non-zero intensity can be obtained by taking into account exactly of terms
related to the interaction between\ the nearest neighbors . Using Eq. (18)
we obtain the analytical expression of the intensities of 6Q coherence
calculated up to second order in ratio of $\frac{\beta }{\alpha }$ and $%
\frac{\delta }{\alpha }$

\begin{equation}
J_{6Q}=-\frac{\left( \frac{\delta }{\alpha }\right) ^{2}}{497664}\left[
84x\left( 2\cos x+\cos 2x\right) +88\sin x-5\left( 32\sin 2x+\sin 4x\right) %
\right] ^{2}  \tag{35}
\end{equation}%
It is interesting that $J_{6Q}$ \ coherence up to second order does not
depend on the interaction between the next-nearest neighbors. The validity
of the time dependence of \ $J_{6Q}$ according to Eq. (35) was tested by
comparison with numerical results. Fig. 3 shows that the prediction of Eq.
(35) gives a reasonable agreement with the simulation data up to $x\leq 7$ ,
which coincides with the estimation using the dipolar coupling constant: $x<%
\frac{\alpha }{\delta }=\allowbreak 8$. The results of our simulation, shown
in Fig. 3, coincide with the dependences obtained in Refs. \cite%
{S.I.Doronin2000,Khitrin}.

\textbf{Conclusion}

In conclusion, the perturbation method was developed which is based on the
expansion of the operator exponent in a perturbation series. It allows us to
apply the perturbation approach to the description of the MQ spin dynamics
in solids. The exact and perturbation analytical expressions were obtained
which describe 0Q and 2Q dynamics in a cluster in the form of{\LARGE \ }a
linear chain{\LARGE \ }of four nuclear spins coupled by DDI in solids. The
calculated 0Q- and 2Q intensities versus the duration of the preparation
period agree with the simulation data \cite{Cho1996}. The obtained
analytical expression for 6Q dynamics in a ring of six spins coupled by DDI
is in a good agreement with the results of numerical simulation. The
developed method can be extended to include various power of the operators%
{\LARGE \ }and applied to the description of a wide range of physical
problems that deal with dynamics of dipolar coupling spins in solids.

\textbf{Acknowledgments}

The authors are grateful to E. B. Fel'dman and I. I. Maximov (Institute of
Problems of Chemical Physics, Chernogolovka) and A. K. Khitrin (Kent State
University, Kent) for useful discussions. This research was supported by a
Grant from the U.S.-Israel Binational Science Foundation (BSF).

Captions for figures.

Figure 1

Fig. 1. Time dependences (in units of $\frac{1}{\alpha }$ ) of the
normalized intensities of 0Q coherence in a{\LARGE \ }linear chain of four
nuclear spins $1/2$ coupled by DDI. Solid line is the exact solution
(Eq.(29)), dotted line is the calculation using Eq.(25), dashed\ line is the
calculation using Eq.(27), open circles present the simulation results of 
\cite{Cho1996}, and down solid triangles are our numerical simulation.

Figure 2

Fig. 2. Time dependences (in units of $\frac{1}{\alpha }$ ) of the
normalized intensities of 2Q coherence in a linear chain of four nuclear
spins $1/2$ coupled by DDI. Solid line is the exact solution (Eq.(30),
dotted line is the calculation using Eq.(28), dashed line is the calculation
using Eq.(26), open circles present the simulation results of \cite{Cho1996}%
, and down solid triangles are our numerical simulation.

Figure 3

Fig.3

Time dependences of the intensities of 6Q coherence in a ring of six spins.
The dashed line is the prediction of Eq. (35) up to second order in\ ratio
of $\frac{\delta }{\alpha }$, and the solid line is the simulation data from 
\cite{S.I.Doronin2000,Khitrin}.

\end{document}